\documentclass[twocolumn,showpacs,secnumarabic,amssymb,nobibnotes,aps,prl]{revtex4-1}
\usepackage{amsmath}
\usepackage{latexsym}
\usepackage{graphicx}
\usepackage{textcomp}
\usepackage{hyperref}
\def\ba{\begin{eqnarray}}
\def\ea{\end{eqnarray}}
\def\be{\begin{equation}}
\def\ee{\end{equation}}
\def\bm{\begin{math}}
\def\me{\end{math}}

\newcommand{\dummy}

\begin{document}

\title{Effects of Density Conservation and Hydrodynamics on Aging in Nonequilibrium Processes}
\author{Suman Majumder and Subir K. Das$^{*}$}
\affiliation{Theoretical Sciences Unit, Jawaharlal Nehru Centre for Advanced Scientific Research,
 Jakkur P.O, Bangalore 560064, India}

\date{\today}

\begin{abstract}
~Aging in kinetics of three different phase transitions, viz., magnetic, binary solid and single component fluid, have been 
studied via Monte Carlo and molecular dynamics simulations in three space dimensions with the objective of identifying the effects 
of order-parameter conservation and hydrodynamics. We observe that the relevant autocorrelations exhibit power-law decay in ferromagnet and binary 
solid but with different exponents. At early time fluid autocorrelation function nicely follows that of binary solid, 
the order parameter being conserved for both of them, as opposed to a ferromagnet. At late time the fluid data crosses 
over to an exponential decay which we identify as a hydrodynamic effect and provide analytical justification for this behavior.

\end{abstract}


\maketitle
~~Understanding properties related to aging \cite{Zannetti} in out of equilibrium systems is of fundamental as well as of practical importance. 
Systems of interest are in abundance \cite{Mathieu,Abou,Costa,Kenning,Berthier,Masri,Bouchbinder,Bergli}, 
starting from biology to cosmology. In the literature of nonequilibrium statistical mechanics, even though the quantities involving 
single time are reasonably well understood \cite{Bray}, those involving multiple time remained difficult. Aging phenomena is 
related to the latter where one expects slower relaxation of older systems. Apart from this obvious 
qualitative fact, understanding of this important phenomena is very poor even for simplest of the systems.
\par
~~In this letter, we present results from the studies of aging kinetics in nonequilibrium systems following three 
distinctly different phase transitions with the objective to understand the effects of order-parameter conservation and 
hydrodynamics on this phenomena. For this purpose we have studied the two-time or autocorrelation function \cite{Zannetti}, 
$C(t,t_w)$, defined as 
\begin{eqnarray}\label{auto_Cr}
 C(t,t_w)=\langle \phi (\vec{r},t)\phi(\vec{r},t_w)\rangle-\langle\phi(\vec{r},t)\rangle\langle\phi(\vec{r},t_w)\rangle,
\end{eqnarray}
where $\phi (\vec{r},t)$ is the relevant space ($\vec{r}$) and time ($t$) dependent order parameter. In 
Eq. (\ref{auto_Cr}), $t$ and $t_w$ are referred to as the observation and waiting times, respectively -- the latter essentially is the 
age of the system.
\par
~~Fisher and Huse (FH) \cite{DFisher}, from the study of spin glass systems, predicted a power-law decay of 
$C(t,t_w)$, in $d$ space dimensions, as 
\begin{eqnarray}\label{auto_PL}
  C(t,t_w) \sim \left( \frac{\ell}{\ell_w} \right)^{-\lambda};~\frac{d}{2} \leq \lambda \leq d,
\end{eqnarray}
where $\ell$ and $\ell_w$ are characteristic lengths of a system at time $t$ and $t_w$, respectively. However, 
not much further information have been obtained either on the value of the exponent or on the general validity of 
a power-law decay, particularly for systems with conserved order-parameter dynamics.
\par
~In this work, from the comparative studies of aging dynamics in $3-d$ systems undergoing ferromagnetic ordering, phase 
separation in a solid binary mixture and that in a vapor-liquid system, we obtain significant general understanding. 
Note that having been quenched from a homogeneous state to a temperature ($T$) below the critical one ($T_c$), these systems 
move towards the new equilibrium state via formation and growth of domains as \cite{Bray}
\begin{eqnarray}\label{powerlaw}
  \ell(t) \sim t^{\alpha}.
\end{eqnarray}
In case of a ferromagnet, where one has nonconserved order-parameter, the exponent $\alpha$ has a value \cite{Bray} $1/2$; 
for phase separating binary solid, for which the order-parameter is a conserved quantity \cite{Bray,LS,Suman_Ising}, $\alpha=1/3$. 
In fluids, however, the entire growth process cannot be described by a single exponent. This complexity is due to the 
influence of hydrodynamics \cite{Siggia,Furukawa,Suman_VL,Shaista}. In this case the early time growth is consistent with the binary solid 
due to diffusive transport. At late time, the exponent crosses over to $\alpha=1$, referred to as the viscous hydrodynamic regime 
and further to the inertial hydrodynamic regime with $\alpha=2/3$. It is crucial to understand the effects of $\alpha$ and thus 
the growth mechanism, on the decay of $C(t,t_w)$.
\par
~To address these issues, we have considered two different models. For the growth dynamics in ferromagnet and solid binary 
mixture, we have studied the nearest neighbor Ising model $H = -J\sum_{<ij>}S_i S_j;~S_{i}=\pm1;~J>0,$ 
prototype for a variety of phase transitions. For a binary (A+B) mixture spin value $S_i=1$ corresponds to an A-particle and 
$-1$ to a B-particle. The kinetics in this model was studied via Monte Carlo (MC) simulations \cite{MCbook}. In the nonconserved 
case we have implemented the Glauber spin-flip kinetics \cite{MCbook} where in a trial 
MC move the sign of a randomly chosen spin was altered and the move was accepted or rejected according to standard 
Metropolis criterion. In case of conserved kinetics, we have used the Kawasaki exchange mechanism \cite{MCbook} where, in 
a trial move, positions of a randomly chosen nearest neighbor pair of spins were interchanged. On the other hand, for the 
vapor-liquid transition, we have carried out molecular dynamics (MD) simulations \cite{Frenkel} on a model where particles 
of equal mass ($m$) interact with each other via \cite{Suman_VL} $u(r=|\vec{r_i}-\vec{r_j}|) = 
U(r)-U(r_c)-(r-r_{c})(dU/dr)_{r=r_{c}},$
where $U(r)=4\epsilon \left[ (\sigma/r )^{12} - (\sigma/r)^{6} \right] $ 
is the standard Lennard-Jones (LJ) potential, with $\epsilon$ and $\sigma$ being respectively the interaction strength and 
particle diameter. Here the cut-off distance $r_c(=2.5\sigma)$ was introduced for faster computation.
\begin{figure}[htb]
  \centering
\includegraphics*[width=0.42\textwidth]{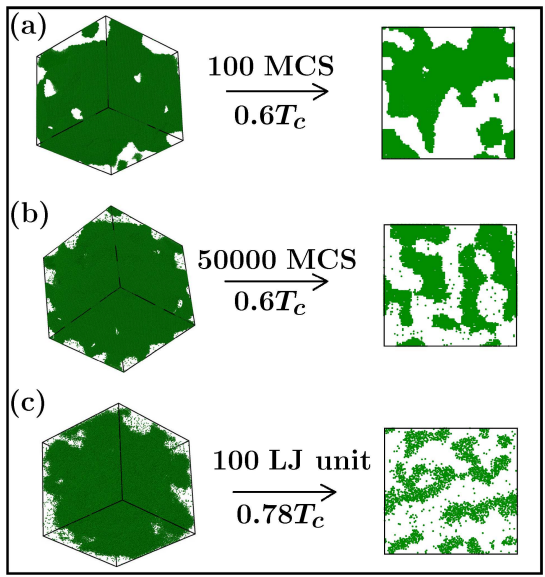}
\caption{\label{fig1} Snapshots during the evolutions of (a) a ferromagnet, (b) a solid binary mixture and (c) a vapor-liquid system, 
at indicated times and temperatures. The left column shows the original $3-d$ snapshots while the right column shows corresponding 
$2-d$ slices. The linear dimension $L$ of the systems are $100$, $100$ and $96$, respectively. In all the snapshots only up spins 
or A-particles are shown.}
 \end{figure}
\par
~As opposed to the MC simulation of Ising model where the spins or particles sit only on sites of a regular lattice system 
(of lattice constant $a$), in the MD simulations particles can change their positions continuously. To control the temperature 
in MD simulations, we have used the Nos\'{e}-Hoover thermostat \cite{Frenkel} that preserves hydrodynamics well. For the Ising model, 
time was measured in units of the Monte Carlo Steps (MCS) with one MCS consisting of $L^3$ trial moves, $L$ being the linear 
dimension of the system. For the MD runs, we have the LJ time unit $t_0=( m\sigma^2/\epsilon)^{1/2}$. In the 
rest of the paper, we set $m$, $\sigma$, $\epsilon$, $J$, $a$ and the Boltzmann constant $k_B$ to unity. Then the LJ unit $t_0$ becomes 
unity as well.  

\par

~In addition to $C(t,t_w)$, $\ell(t)$, that will often be used, was obtained from the first moment of the domain 
size distribution function $P(\ell_d,t)$ as $\ell(t)= \int d\ell_d \ell_d P(\ell_d,t),$ 
where $\ell_d$ is the distance between two successive domain boundaries in $x-$, $y-$, or $z-$ directions. 
All our results are obtained with periodic boundary conditions. For the Ising model we have chosen a simple cubic lattice. 
For the analysis of LJ system results, the continuous original configurations were mapped to a 
simple cubic lattice \cite{Suman_VL}. In this procedure, every particle 
was moved to the nearest lattice site, following which if a site is occupied by a particle it was  assigned a spin value $+1$, 
otherwise $-1$. Note that in Eq. (\ref{auto_Cr}) $\phi$ corresponds to these spin values. 
Quantitative results are presented after averaging over multiple initial configurations. 
Before presenting the results, we mention that for the Ising model \cite{MCbook} $T_c=4.51$ and for the LJ 
system it is \cite{Sutapa} $\simeq 0.9$. In all the cases, quenching was done along the critical composition or density ($\rho$). 
For Ising model, of course, this corresponds to a $50:50$ composition of up and down spins, while for the LJ model we have 
\cite{Suman_VL,Sutapa} $\rho_c \simeq 0.3$. 

\par
~In Fig. \ref{fig1}, we show snapshots from the evolutions of all three systems, starting from homogeneous 
initial configurations. The temperatures of quench in each of the cases are mentioned on the figure. The frames 
on the left are original $3-d$ configurations. For better judgement of the pattern, on the right hand side frames we show 
$2-d$ slices of the systems. It appears that there is significant difference in morphology for the conserved and 
nonconserved dynamics \cite{Bray}. Also, if the dynamics is conserved, effect of hydrodynamics does not, 
at least, bring visibly different features in the pattern as is clear from the snapshots from binary solid and 
vapor-liquid systems.
\begin{figure}[htb]
  \centering
\includegraphics*[width=0.45\textwidth]{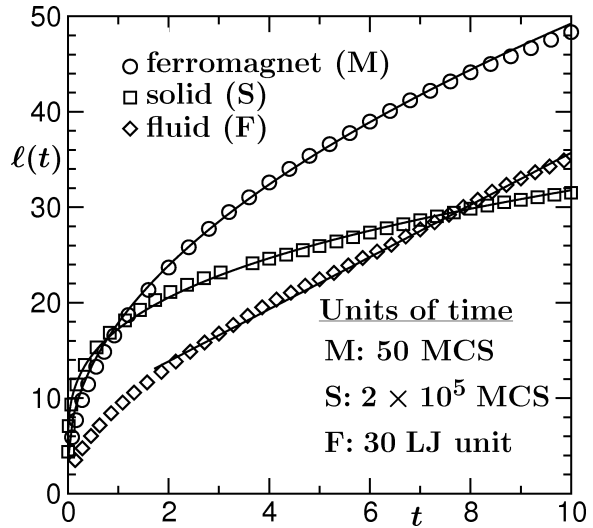}
\caption{\label{fig2} Plots of average domain size, $\ell(t)$, vs time, for various coarsening systems. The continuous 
lines represent expected functional behaviors. The units of time for various systems are indicated on the figure.}
 \end{figure}
\par
~The growth dynamics is compared, for all the systems, in Fig. \ref{fig2} where we plot $\ell(t)$ as a function of $t$. 
The continuous lines in this figure are fits to the expected theoretical exponents. For the Ising model, it is clearly 
seen that the data are consistent with exponents $\alpha=1/3$ and $1/2$ for the conserved and nonconserved dynamics 
respectively. For the LJ model, however, after a brief period of slow growth, there is a crossover to a regime where 
the simulation data are nicely consistent with a linear behavior corresponding to viscous hydrodynamic growth \cite{Suman_VL}. 
We do not aim for a further crossover to the inertial hydrodynamic regime due to lack of computational resources. Next we move 
to the central objective of the paper. 
\par
\begin{figure}[htb]
  \centering
\includegraphics*[width=0.42\textwidth]{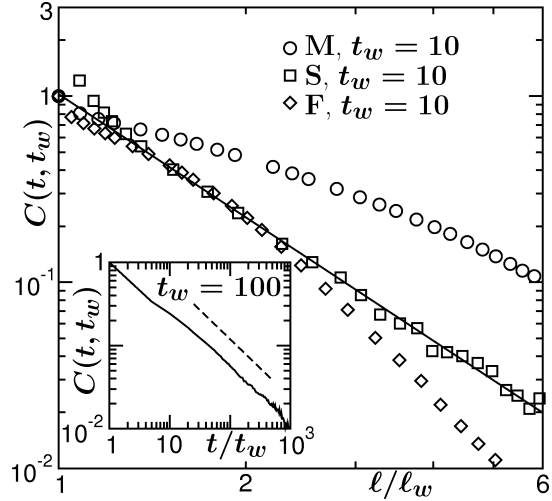}
\caption{\label{fig3} Log-log plots of $C(t,t_w)$ vs $\ell/\ell_w$ for all the three systems. The solid line corresponds 
to a power-law decay with an exponent $\lambda=2.2$. The ordinate of the binary mixture data was multiplied by a number to obtain overlap 
with the fluid data at appropriate region. Inset shows $C(t,t_w)$ vs $t/t_w$ for the solid mixture. There the 
dashed line has a power-law exponent $-0.7$.}
 \end{figure}

~In Fig. \ref{fig3} we show the plots of $C(t,t_w)$ vs $\ell/\ell_w$ for all the systems. A double-log scale is used and the values 
of $t_w$ in each of the cases are mentioned on the figure. It is seen that the data for the solid binary mixture is consistent with 
a power-law decay starting from very small value of the abscissa variable till the end of the available simulation results. The continuous 
line there has an exponent $\lambda=2.2$ with which this set of data are nicely consistent. The results for ferromagnet or vapor-liquid 
system does not show linear look over the whole range. However, the fluid data are consistent with the solid mixture result 
at the beginning. This is due to the fact that hydrodynamics becomes important only at large length scales as already seen in 
Fig. \ref{fig2}. Around the value of $\ell(t)$ from where we have seen a linear behavior for the LJ system in Fig. \ref{fig2}, 
we observe a deviation from the power-law in this figure. One can also ask, if the continuous curvature of the ferromagnetic 
data is also indicative of a non-power-law decay? Before moving on to answer these questions, in the inset of Fig. \ref{fig3} we 
show  $C(t,t_w)$ vs $t/t_w$ for the binary solid only. Again the data look very linear, after a brief initial period, 
on a double log scale and are consistent 
with a power-law exponent $\simeq -0.7$. This indirectly confirms that for diffusive kinetics with conserved order parameter 
$\alpha$ is $1/3$ with which, of course, early time fluid results are consistent. 
The value $\lambda=2.2$ is, of course, consistent with the bound predicted by FH \cite{DFisher,Marko}. 
However, this is in disagreement with similar studies \cite{Desai} via Cahn Hilliard equation \cite{Bray}.

\par
\begin{figure}[htb]
  \centering
\includegraphics*[width=0.42\textwidth]{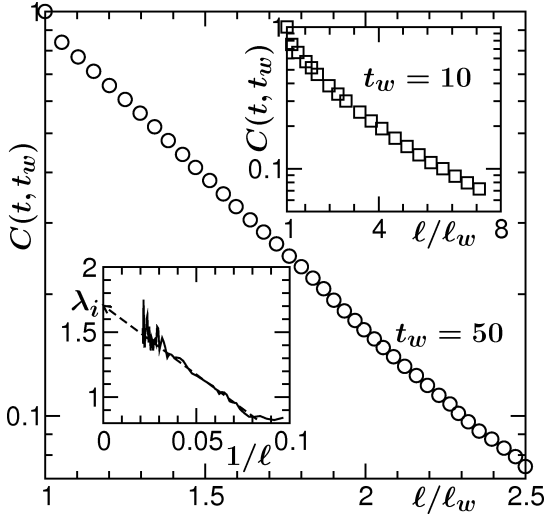}
\caption{\label{fig4} Log-linear plot of $C(t,t_w)$ vs $\ell/\ell_w$ for the vapor-liquid system. 
The upper inset shows the corresponding plot for a ferromagnet. The lower inset shows the instantaneous exponent 
$\lambda_i$ for the ferromagnetic system, as a function of $1/\ell$. The solid line with an arrow is a guide to the eye.}
 \end{figure}

~Next, in the main frame of Fig. \ref{fig4} we show log-linear plots of $C(t,t_w)$ vs $\ell/\ell_w$ for the vapor-liquid transition. 
Note here that we have chosen a value of $t_w$ such that the system is in the hydrodynamic regime. The minimum 
value of $t_w$ needed for this can be read out from Fig. \ref{fig2}. 
This data clearly looks linear on this plot which confirms exponential decay 
of the autocorrelation function. The ferromagnet data, presented in the upper inset of Fig. \ref{fig4}, 
however, is inconsistent with it. Note here that in all the cases we have 
experimented with various values of $t_w$ for general understanding of scaling with respect to $t_w$ or $\ell_w$ 
and for the sake of brevity presented only the representative ones. 
\par
~In view of the other expectation that the ferromagnetic data follow a power-law behavior with time or length dependent correction, in 
the lower inset of Fig. \ref{fig4} we present the instantaneous exponent \cite{Huse} $\lambda_i= -d \ln C(t,t_w)/d \ln \ell,$ 
as a function of $1/\ell$. It appears that $\lambda_i$ has a linear dependence on $1/\ell$, 
for a significant period of time. The minor deviation at late time, we confirmed, is 
a finite-size effect. If we neglect the part affected by finite-size effects, the data converge to a value $\simeq 1.7$, predicted 
by theoretical calculations of Liu and Mazenko \cite{Liu}. Thus there is no universality involving conserved and nonconserved dynamics, 
even though in both cases $C(t,t_w)$ follow power-law decay. 
\par
~Finally, we come to the understanding of the exponential decay, that was also recently observed in liquid-liquid transition \cite{Lippiello}, 
for the vapor-liquid transition in the hydrodynamic regime. To accomplish that we start with the order-parameter update equation of 
model H \cite{Bray}
\begin{eqnarray}\label{model_H}
 \frac{ \partial \phi}{\partial t} + \vec{v}. \nabla \phi = D \nabla^2 \mu,
\end{eqnarray}
where $\vec{v}$ is the advection field, $D$ is a diffusion constant and $\mu$ is the chemical potential. 
With the understanding that in the fast hydrodynamic regime, 
contribution from diffusion is negligible, we neglect the term on the right hand side. From the definition of the autocorrelation function in 
Eq. (\ref{auto_Cr}), it is clear that our task is to show that the order-parameter changes exponentially fast. 
Here we make an assumption that this exponential decay is due to fast interfacial motion. Noting that for the viscous hydrodynamic growth, 
$v(=\ell/t)$ is constant ($C$) and in the interfacial region $\nabla \phi \rightarrow 2\phi/w $, $w$ being the interfacial width, we obtain
$ d \phi/dt =-K \phi,$ where $K(=2C/w)$ is a constant. This provides $\phi \sim \exp \left( -2 \ell/ w \right)$.
\par
~In conclusion, we have presented results for aging during the nonequilibrium evolutions in various systems -- 
ferromagnet, solid binary mixture and vapor-liquid system -- following quench from high temperatures below the critical ones. 
The two-time correlation function \cite{Zannetti} $C(t,t_w)$ has been used as a probe for this study. The Ising 
model with conserved and nonconserved order-parameter dynamics were used to study the binary solid and ferromagnetic systems, respectively, 
while an LJ model \cite{Suman_VL} was used for the study of fluid. 
\par
~It is observed that in absence of hydrodynamics, in all the cases the autocorrelation decays in a power-law fashion. The exponent 
for the conserved order-parameter dynamics deviates significantly from the nonconserved one, but both of them 
follow the bounds predicted by Fisher and Huse \cite{DFisher}. Interestingly, for the nonconserved case, there is significant curvature 
dependent correction to the exponent. At late time, in fluid, there is a crossover from power-law 
to an exponential behavior which we understood via analytical argument. 

\section*{Acknowledgment}\label{ack}
~SKD acknowledges important discussion with M. Zannetti, S. Puri, F. Corberi and E. Lippiello. Both of us acknowledge financial support 
from the Department of Science and Technology, India, via grant no. SR/S2/RJN-13/2009. SM is grateful to Council of Scientific and 
Industrial Research, India, for research fellowship. 
\vskip 0.5cm
${*}$ das@jncasr.ac.in
 

\begin{thebibliography}{100}
 \bibitem{Zannetti} M. Zannetti in \textit{Kinetics of Phase Transitions}, edited by 
 S. Puri and V. Wadhawan (2009).
\bibitem{Mathieu} R. Mathieu, P. Norblad, D.N.H. Nam, N.X. Phue and N.V. Khiem, 
 Phys. Rev. B \textbf{63}, 174405 (2001).
\bibitem{Abou} B. Abou and F. Gallet, Phys. Rev. Lett. \textbf{93}, 160603 (2004).
 \bibitem{Costa} M. Costa, A.L. Goldberger and C.-K. Peng, Phys. Rev. Lett. \textbf{95}, 198102 (2005).
\bibitem{Kenning} G.G. Kenning, G.F. Rodriguez and R. Orbach, Phys. Rev. Lett. \textbf{97}, 057201 (2006).
\bibitem{Berthier} L. Berthier, Phys. Rev. Lett. \textbf{98}, 220601 (2007).
\bibitem{Masri} D.E. Masri, L. Berthier and L. Cipelletti, Phys. Rev. E \textbf{82}, 031503 (2007).
\bibitem{Bouchbinder} E. Bouchbinder and J.S. Langer, Phys. Rev. E \textbf{83}, 061503 (20011).
\bibitem{Bergli} J. Bergli and Y.M. Galperin, Phys. Rev. B \textbf{85}, 214202 (20012).
\bibitem{Bray} A.J. Bray, Adv. Phys. \textbf{51}, 481 (2002).
\bibitem{DFisher} D.S. Fisher and D.A. Huse, Phys. Rev. B, \textbf{38}, 373 (1989).
\bibitem{LS} I.M. Lifshitz and V.V. Slyozov, J. Phys. Chem. Solids \textbf{19}, 35 (1961).
\bibitem{Suman_Ising} S. Majumder and S.K. Das,  Phys. Rev. E \textbf{81}, 050102 (2010).
\bibitem{Siggia} E.D. Siggia, Phys. Rev. A \textbf{20}, 595 (1979).
\bibitem{Furukawa} H. Furukawa, Phys. Rev. A \textbf{31}, 1103 (1985).
\bibitem{Suman_VL} S. Majumder and S.K. Das,  Europhys. Lett. \textbf{95}, 46002 (2011).
\bibitem{Shaista} S. Ahmad, S.K. Das and S. Puri, Phys. Rev. E \textbf{82}, 040107 (2010).
\bibitem{MCbook} D.P. Landau and K. Binder, \textit{A Guide to Monte Carlo Simulations in Statistical Physics},
 Cambridge University Press, 3rd Edition (2009).
\bibitem{Frenkel} D. Frenkel and B. Smit, \textit{Understanding Molecular Simulations:
 From Algorithm to Applications} (Academic Press, San Diego, 2002).
\bibitem{Sutapa} S. Roy and S.K. Das, Phys. Rev. E \textbf{85}, 050602 (2012).
\bibitem{Marko} J.F. Marko and G.T. Barkema, Phys. Rev. E \textbf{52}, 2522 (1995).
\bibitem{Desai} C. Yeung, M. Rao and R.C. Desai, Phys. Rev. E \textbf{53}, 3073 (1996).
\bibitem{Huse} D.A. Huse, Phys. Rev. B, \textbf{40}, 304 (1989).
\bibitem{Liu} F. Liu and G.F. Mazenko, Phys. Rev. B \textbf{44}, 9185 (1991).
\bibitem{Lippiello} S. Ahmad, F. Corberi, S.K. Das, E. Lippiello, S. Puri and M. Zannetti, Phys. Rev. E \textbf{86}, 061129 (2012).
%






%
 \end{thebibliography}
\end{document}